\documentclass[preprint,showpacs,preprintnumbers,amsmath,amssymb]{revtex4}
\usepackage{bbm}

\usepackage{graphicx}
\usepackage{dcolumn}
\usepackage{bm}
\usepackage{amsmath}

\begin{document}


\title{Effective generation of Ising interaction and cluster states in coupled microcavities}

\author{Pengbo Li$^{1}$}
\author{ Qihuang Gong$^{1}$}
\author{ Guangcan Guo$^{1,2}$}
\affiliation{$^{1}$State Key Laboratory for Mesoscopic Physics,
Department of Physics, Peking University, Beijing 100871, China\\
$^{2}$Key Laboratory of Quantum Information, University of Science
and Technology of China, Hefei 230026, China}

\date{\today}

\begin{abstract}
We propose a scheme for realizing the Ising spin-spin interaction
and atomic cluster states utilizing trapped atoms in coupled
microcavities. It is shown that the atoms can interact with each
other via the exchange of virtual photons of the cavities. Through
suitably tuning the parameters, an effective Ising spin-spin
interaction can be generated in this optical system, which is used
to produce the cluster states. This scheme does not need the
preparation of initial states of atoms and cavity modes, and is
insensitive to cavity decay.
\end{abstract}

 \maketitle

Strongly correlated many-particle systems have been extensively
explored in condensed matter physics, cold
atoms\cite{RMP-78-179,ann.phy}, and recently in optical system such
as coupled microcavity
lattices\cite{Nature-physics-2-849,Nature-physics-2-856,Prl-99-160501,Prl-99-103601,Prl-99-183602,pra-66-031805,eprint1}.
Compared to other strongly correlated many particle systems, an
optical system has the advantage of easily addressing individual
lattice sites with optical lasers. Because of the size and
separation of the microcavities, arbitrary lattice geometries can be
arranged in this system. Therefore, it offers the ability to
experimentally observe quantum-many-body phenomena and realize
quantum information processing. Various technologies have been
employed in this field, including fiber coupled toroidal
microcavities\cite{nature-421}, arrays of defects in photonic band
gap materials\cite{Sci308}, and superconducting qubits coupled
through microwave stripline resonators\cite{nature-431}.

On the other hand, interacting qubits are particularly important in
quantum information processing. Lattices of interacting spins or
qubits can be utilized to generate highly entangled states, such as
cluster states\cite{Prl-86-910,eprint2}. It has been shown that this
class of entangled states are much more entangled than the GHZ
states, thus have a high persistence of
entanglement\cite{Prl-86-910,eprint2}. The experimental
demonstration of the violation of Bell's inequality for cluster
states has been reported\cite{Prl-95-020403}. Moreover, cluster
states together with local measurements are the resource for one-way
quantum computation\cite{Prl-86-5188,nature-434}. In the context of
cavity QED\cite{Sci298,RMP-73-565,Kimble}, several proposals have
been presented for producing the atomic cluster
states\cite{Prl-95-160501,pra-72-034304}. However, the scalable
implementation in many-particle case is difficulty. Recently in Ref.
\cite{Prl-99-160501} a very novel scheme has been proposed to
simulate the dynamics of an effective anisotropic Heisenberg
spin-$1/2$ chains and generate atomic cluster states in coupled
cavities. However, there is no report of the realization of Ising
interaction and cluster states in a simple enough atom-cavity system
that does not require the complicated atomic level structure and
driving laser configurations.

In this paper we present a scheme for realizing the Ising spin-spin
interaction and atomic cluster states utilizing trapped two-level
atoms in coupled microcavities arranged in an array. We show that
the atoms can interact with each other via the exchange of virtual
photons of the cavities. Through suitably tuning the parameters, an
effective Ising spin-spin interaction can be generated in this
optical system. We discuss how to use this Ising interaction to
produce the cluster states of atoms. This scheme does not need the
preparation of initial states of atoms and cavity modes, and is
insensitive to cavity decay. With presently available experimental
setups in cavity QED, the implementation of this scheme is feasible.

As sketched in Fig. 1, an array of microcavities are coupled via the
exchange of photons with identical two-level atom in each cavity.
The ground state of each atom is labeled as $\vert g_j\rangle$, and
the excited state as $\vert e_j\rangle$, where the index $j$ counts
the cavities. The cavity mode (frequency $\omega_c$) couples to the
transition $\vert g\rangle\leftrightarrow \vert e\rangle$
(transition frequency $\omega_0$) with coupling constants $g$.
Furthermore, an external strong classical field (frequency
$\omega_L$) drives the same transition with Rabi frequencies
$\Omega$\cite{Prl-90-027903}. For simplicity, we only consider the
one-dimensional array. One can generalize to the higher dimensions
straightforwardly. In the rotating-wave approximation, the
associated Hamiltonian reads (let $\hbar=1$)
\begin{eqnarray}
\label{h1}
H&=&H_a+H_c+H_{ac},\nonumber\\
H_a&=&\sum^N_{j=1}\omega_0S^+_jS^-_j\nonumber\\
H_c&=&\omega_c\sum^N_{j=1}\hat{a}^\dag_j\hat{a}_j+J_c\sum^N_{j=1}(\hat{a}^\dag_j\hat{a}_{j+1}+\hat{a}_j\hat{a}^\dag_{j+1}),\nonumber\\
H_{ac}&=&\sum^N_{j=1}[g(\hat{a}^\dag_jS_j^-+\hat{a}_jS_j^+)+\Omega(S_j^+e^{-i\omega_Lt}+S_j^-e^{i\omega_Lt})],\nonumber\\
\end{eqnarray}
where $S_j^+=\vert e_j\rangle\langle g_j\vert$,
$S_j^-=(S_j^+)^\dag$, $\hat{a}_j$ is annihilation operator for the
photon in cavity $j$, and $J_c$ is the hopping rate of photons
between neighboring cavities. In the complete Hamiltonian, $H_a$
describes the free Hamiltonian for atoms; $H_c$ describes the
Hamiltonian for photons in each cavity modes, with photon hopping
between neighboring cavities; finally $H_{ac}$ is the Hamiltonian
that describes the interaction between atoms and the cavities as
well as the strong driving by the classical fields. We consider the
periodic boundary conditions, then $H_c$ can be diagonalized through
the Fourier transformation. For convenience we introduce the
notation $\textbf{J}=(uj,0,0)$ to denote the position of the $j$th
site where $u$ is the length of the one-dimensional crystal cell and
in the following chosen as unit for simplicity. Then we obtain
$H_c=\sum_k\omega_ka_k^\dag a_k$, where $\omega_k=\omega_c+2J_c\cos
k$.

The Hamiltonian of Eq. (\ref{h1}) can be changed to a reference
frame rotating with the driving field frequency $\omega_L$,
\begin{eqnarray}
H=&&\Delta\sum^N_{j=1}S^+_jS^-_j-\sum_k\delta_ka_k^\dag
a_k+\sum_{j=1}^N\Omega(S_j^++S_j^-)\nonumber\\
 &&+\sum_{j=1}^N[gS_j^-\sum_k
a_k^\dag e^{i\textbf{k}\cdot\textbf{J}}+gS_j^+\sum_k a_k
e^{-i\textbf{k}\cdot\textbf{J}}],
\end{eqnarray}
where $\Delta=\omega_0-\omega_L$, and $\delta_k=\omega_L-\omega_k$.
In the following we assume $\omega_0=\omega_L$ for simplicity. We
now switch to a new atomic basis
$\vert\downarrow_j\rangle=\frac{1}{\sqrt{2}}(\vert g_j\rangle+\vert
e_j\rangle)$ and $\vert\uparrow_j\rangle=\frac{1}{\sqrt{2}}(\vert
g_j\rangle-\vert e_j\rangle)$, then can rewrite $H$ as
\begin{eqnarray}
H=&&-\sum_k\delta_ka_k^\dag
a_k+\sum_{j=1}^N\Omega\sigma_j^z\nonumber\\
 &&+\sum_{j=1}^N[g(\frac{1}{2}\sigma_j^z+\frac{1}{2}\sigma_j^+-\frac{1}{2}\sigma_j^-)\sum_k
a_k^\dag e^{i\textbf{k}\cdot\textbf{J}}\nonumber\\
&&+g(\frac{1}{2}\sigma_j^z+\frac{1}{2}\sigma_j^--\frac{1}{2}\sigma_j^+)\sum_k
a_k e^{-i\textbf{k}\cdot\textbf{J}}].
\end{eqnarray}
where $\sigma^z_j=\vert \downarrow_j\rangle\langle
\downarrow_j\vert-\vert \uparrow_j\rangle\langle \uparrow_j\vert$,
$\sigma_j^+=\vert \uparrow_j\rangle\langle \downarrow_j\vert$, and
$\sigma_j^-=(\sigma_j^+)^\dag$. In the interaction picture with
respect to $H_0=-\sum_k\delta_ka_k^\dag
a_k+\sum_{j=1}^N\Omega\sigma_j^z$, we have the following interaction
Hamiltonian\cite{Prl-90-027903}
\begin{eqnarray}
H_I=&&\sum_{j=1}^N[g(\frac{1}{2}\sigma_j^z+\frac{1}{2}\sigma_j^+e^{-i\Omega
t}-\frac{1}{2}\sigma_j^-e^{i\Omega t})\sum_k
a_k^\dag e^{i\textbf{k}\cdot\textbf{J}-i\delta_kt}\nonumber\\
&&+g(\frac{1}{2}\sigma_j^z+\frac{1}{2}\sigma_j^-e^{i\Omega
t}-\frac{1}{2}\sigma_j^+e^{-i\Omega t})\sum_k a_k
e^{-i\textbf{k}\cdot\textbf{J}+i\delta_kt}].\nonumber\\
\end{eqnarray}
In the strong driving regime $\Omega\gg g, \delta_k$ (for all $k$),
we can realize a rotating-wave approximation and neglect the fast
oscillating terms. Then $H_I$ reduces to
\begin{eqnarray}
H_I&=&\sum_{j=1}^N\frac{1}{2}g\sigma_j^z(\sum_k a_k^\dag
e^{i\textbf{k}\cdot\textbf{J}-i\delta_kt} +\sum_k a_k
e^{-i\textbf{k}\cdot\textbf{J}+i\delta_kt}).\nonumber\\
\end{eqnarray}

To further reduce the model, we assume $\delta_k\gg g$ (for all
$k$). Then there is no energy exchange between the atomic system and
the cavities. We can adiabatically eliminate the photons from the
description\cite{James}. We consider the terms up to second order in
the effective Hamiltonian and drop the fast oscillating terms. Then
we obtain the following effective Hamiltonian describing the Ising
type spin-spin interaction in the optical system
\begin{equation}
\label{Ising}
H_I=\sum_{j=1}^NJ_z\sigma^z_j\sigma^z_{j+1},
\end{equation}
where $J_z=\sum_k\frac{g^2e^{ik}}{2\delta_k}$. The parameters $J_z$
can be tuned by varying coupling strength $g$ and detuning
$\delta_k$. The evolution operator for the system is given by
\begin{equation}
\label{U(t)}
U(t)=e^{-iH_It}=e^{-i[\sum^N_{j=1}J_z\sigma^z_j\sigma^z_{j+1}]t}.
\end{equation}
It has been shown that Ising interaction can be utilized to generate
a new class of multipartite entanglement, the so called cluster
states\cite{Prl-86-910}. In the following discussions we will use
this Hamiltonian to produce the cluster states.

In order to reduce notations we will not use the Ising interaction
in Eq. (\ref{Ising}) but rather the phase gate
\begin{eqnarray}
\label{PG} U_p(t)=e^{-iH'_It}\quad\mbox{with}\quad
H'_I=-\sum^N_{j=1}J_zt\frac{1+\sigma^z_j}{2}\frac{1-\sigma^z_{j+1}}{2}
\end{eqnarray}
as the elementary two-qubit interaction between neighboring
atoms\cite{Prl-86-910,eprint2}. The equivalence between the quantum
Ising interaction $H_I$ in Eq. (\ref{Ising}) and $H'_I$ in Eq.
(\ref{PG}) can be seen from the following discussion. From
\begin{eqnarray}
H'_I=&&-\sum^N_{j=1}J_zt\frac{1+\sigma^z_j}{2}\frac{1-\sigma^z_{j+1}}{2}\nonumber\\
=&&-\frac{1}{4}\sum^N_{j=1}J_zt(1+\sigma_j^z-\sigma_{j+1}^z-\sigma_j^z\sigma_{j+1}^z)
\end{eqnarray}
we find
\begin{eqnarray}
U_p(t)=&&e^{\frac{i}{4}\sum^N_{j=1}J_zt}e^{\frac{i}{4}\sum^N_{j=1}J_zt\sigma_j^z}\nonumber\\
&&\times e^{-\frac{i}{4}\sum^N_{j=1}J_zt\sigma_{j+1}^z}
e^{-\frac{i}{4}\sum^N_{j=1}J_zt\sigma_j^z\sigma_{j+1}^z}.
\end{eqnarray}
Therefore, the phase gate corresponds to the Ising interaction up to
some additional $\frac{J_zt}{4}$-rotations around the $z$-axes at
each qubit\cite{eprint2}. The entanglement properties are determined
by the pure Ising interaction, which are not changed by the
$z$-rotations.

We first consider the case of two coupled cavities. We denote $\vert
+_j\rangle\equiv\frac{1}{\sqrt{2}}(\vert \downarrow_j\rangle +\vert
\uparrow_j\rangle)=\vert g_j\rangle$, and $\vert
-_j\rangle\equiv\frac{1}{\sqrt{2}}(\vert \downarrow_j\rangle -\vert
\uparrow_j\rangle)=\vert e_j\rangle(j=1,2)$ for atomic states.
Assume that initially the atom in each cavity is prepared in state
$\vert +_j\rangle=\vert g_j\rangle$. After an interaction time of
$t=\pi/J_z$, we find that
\begin{eqnarray}
U_p^{1,2}(\pi/J_z)=P^1_{z,-}\otimes\mathbbm{l}^2+P^1_{z,+}\otimes\sigma^z_2,
\end{eqnarray}
where $P^j_{z,\pm}=\frac{1\pm\sigma^z_j}{2}(j=1,2)$. Then the state
evolution of the system is given by
\begin{eqnarray}
U_p(\pi/J_z)\vert +_1\rangle\vert
+_2\rangle=\frac{1}{\sqrt{2}}(\vert \downarrow_1\rangle\vert
-_2\rangle+\vert
\uparrow_1\rangle\vert+_2\rangle)\nonumber\\
=\frac{1}{2}(\vert \downarrow_1\rangle\sigma_2^z+\vert
\uparrow_1\rangle)(\vert \downarrow_2\rangle+\vert
\uparrow_2\rangle).
\end{eqnarray}
This state is a maximally entangled state. Up to a local unitary
transformation on qubit 2, we can write it in the standard form.

We now turn to the case of many coupled microcavities. Initially all
the atoms are prepared in states $\vert +_j\rangle=\vert g_j\rangle
(j=1,2,...,N)$. We choose $t=\pi/J_z$, then the produced state can
be written in the following compact form\cite{Prl-86-910,eprint2}
\begin{eqnarray}
\vert \psi_N\rangle=\frac{1}{2^N}\bigotimes^N_{j=1}(\vert
\downarrow_j\rangle\sigma_{j+1}^z+\vert \uparrow_j\rangle).
\end{eqnarray}
Then for $N=3,4$, one can obtain
\begin{eqnarray}
\vert \psi_3\rangle&=&\frac{1}{\sqrt{2}}(\vert
\downarrow_1\rangle\vert \downarrow_2\rangle\vert
\downarrow_3\rangle+\vert \uparrow_1\rangle\vert
\uparrow_2\rangle\vert \uparrow_3\rangle)_{l.u.}\nonumber\\
\vert \psi_4\rangle&=&\frac{1}{2}(\vert \downarrow_1\rangle\vert
\downarrow_2\rangle\vert \downarrow_3\rangle\vert
\downarrow_4\rangle+\vert \downarrow_1\rangle\vert
\downarrow_2\rangle\vert \uparrow_3\rangle\vert
\uparrow_4\rangle\nonumber\\
&&+\vert \uparrow_1\rangle\vert \uparrow_2\rangle\vert
\downarrow_3\rangle\vert \downarrow_4\rangle-\vert
\uparrow_1\rangle\vert \uparrow_2\rangle\vert \uparrow_3\rangle\vert
\uparrow_4\rangle)_{l.u.}.\nonumber\\
\end{eqnarray}
Where ``l.u.'' indicates the equality holds up to a local unitary
transformation on one or more of the
qubits\cite{Prl-86-910,eprint2}. $\vert \psi_3\rangle$ corresponds
to a GHZ state of three qubits, but $\vert \psi_4\rangle$ is not
equivalent to a 4-qubit GHZ state. The state $\vert \psi_4\rangle$
is the 4-qubit cluster state.

It is necessary to verify the approximations by numerics. We
numerically simulate the dynamics generated by the full Hamiltonian
$H$ and compare it with the results generated by the effective model
(\ref{Ising}). As an example here we only consider the case of two
atoms in two cavities. Initially the atoms stay in the state $\vert
g\rangle_1\vert g\rangle_2$, and the cavity mode in the vacuum. In
Fig. 2(a) we display the occupation probability $p_{g_1g_2}$ of
system in the state $\vert g_1\rangle\vert g_2\rangle$, and the
occupation of the photon number occupation $p_{N_1}=\langle
\hat{a}^\dag_1\hat{a}_1\rangle$ calculated both from the full
Hamiltonian and the effective model. Fig. 2(b) shows the von Neumann
entropy of the reduced density matrix of one spin, i.e.,
$E_{vN}=-\mbox{Tr}(\rho_1\log_2\rho_1(t))$. A maximally entangled
state for the atoms occurs for $t=(2n+1)\pi/4J_z$ for any integer
$n$. We choose the parameters as $\Omega=50$ GHZ, $g=0.1$ GHZ,
$J_c=0.02$ GHZ and $\omega_c-\omega_L=1$ GHZ. It can be seen that
the effective model can describe the dynamics very well provided
that the parameters are appropriately chosen. The occupations of the
photon number are always smaller than 0.01. Discrepancies between
the numerical results for the full Hamiltonian and the effective
Hamiltonian are due to the higher order terms for the detunings and
Rabi frequencies. However, these discrepancies are below $2\%$ with
respect to the results from the full Hamiltonian.

We now consider some experimental matters. For experimental
implementation, the parameter of the effective Hamiltonian $J_z$
must be much larger than decay rates of cavity and atomic excited
states. The cavity decay has neglectable effect on this scheme. We
have only to consider the effect of the decay of atomic excited
states. This proposal requires that the life time of the atomic
excited states be longer than the time needed to complete the whole
procedure. For potential atomic system, Rydberg atoms are good
candidate. Promising candidates for microcavities are photonic
bandgap cavities\cite{Sci308}, and toroidal or spherical
microcavities coupled via tapered optical fibers\cite{nature-421}.

In summary, we have proposed a scheme for realizing the Ising
spin-spin interaction and atomic cluster states utilizing trapped
atoms in coupled microcavities. It is shown that the atoms can
interact with each other via the exchange of virtual photons of the
cavities. Through suitably tuning the parameters, an effective Ising
spin-spin interaction can be generated in this optical system. We
discuss how to use this Ising interaction to produce the cluster
states of atoms. This scheme does not need the preparation of
initial states of atoms and cavity modes, and is insensitive to
cavity decay. With presently available experimental setups in cavity
QED, it may be implemented.

This work was supported by the National Natural Science Foundation
of China under Grants Nos. 10674009, 10334010, 10521002, 10434020
and National Key Basic Research Program No.2006CB921601. Pengbo Li
acknowledges the quite useful discussions with Hongyan Li.


\newpage
\begin{figure}[h]
\centering
\includegraphics[bb=120 316 484 631,totalheight=2.5in,clip]{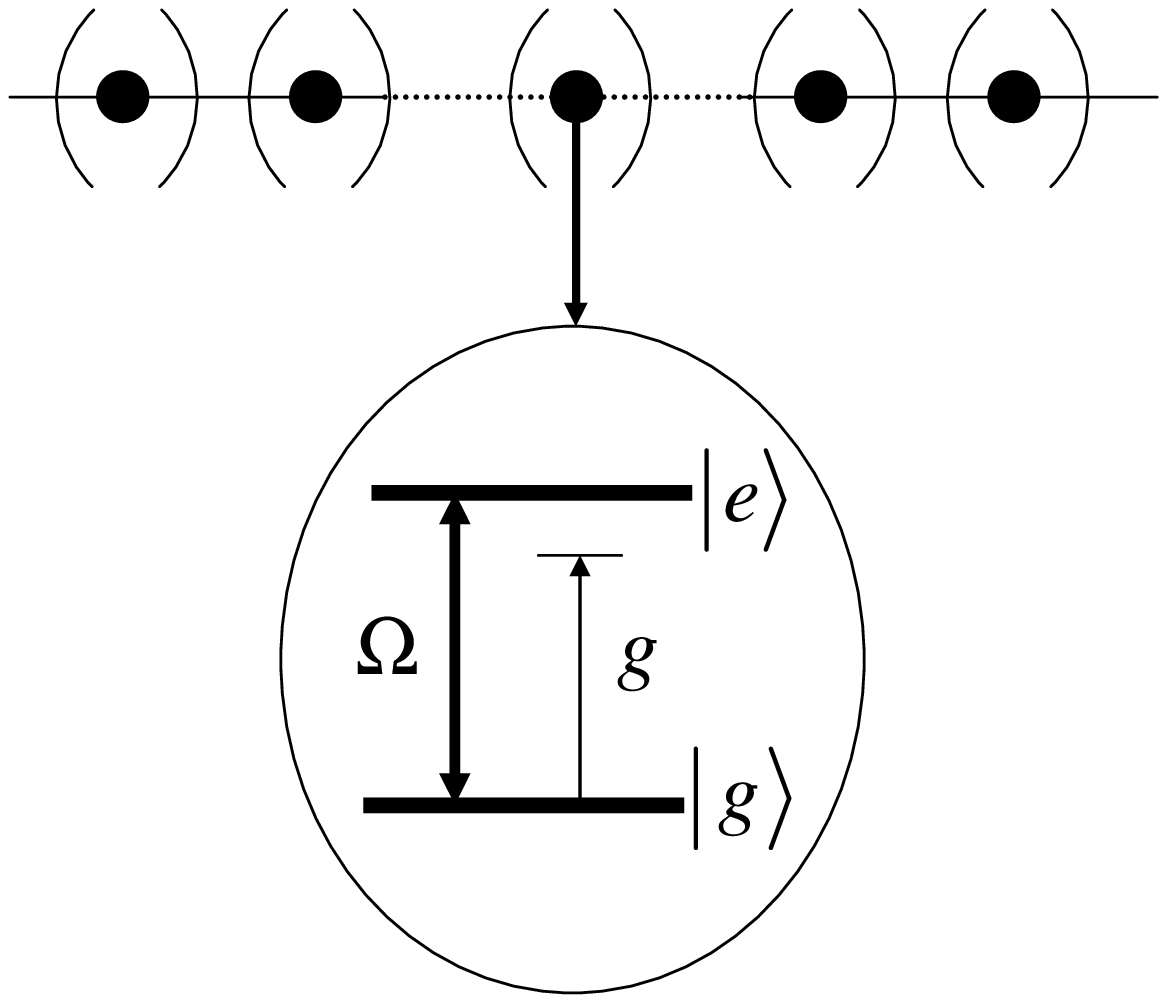}
\caption{An array of microcavities as described in the scheme.
Photon hopping occurs because of the overlap of cavity modes of
adjacent resonators. Two-level atom in each cavity is driven by
external strong field.}
\end{figure}

\newpage
\begin{figure}[h]
\centering
\includegraphics[bb=50 36 576 334,totalheight=2.5in,clip]{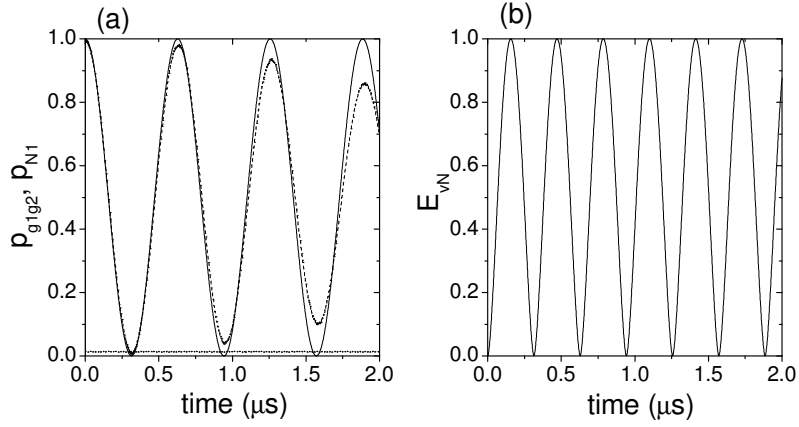}
\caption{ (a) The occupation probability $p_{g_1g_2}$ of the system
in the state $\vert g_1\rangle\vert g_2\rangle$(solid line
represents results from the effective model, and dash line
represents results from the full Hamiltonian), and the photon number
occupation $p_{N_1}=\langle \hat{a}^\dag_1\hat{a}_1\rangle$. (b) The
von Neumann entropy $E_{vN}$ of the reduced density matrix of 1
effective spin. Parameters are chosen as $\Omega=50$ GHZ, $g=0.1$
GHZ, $J_c=0.02$ GHZ and $\omega_c-\omega_L=1$ GHZ.}
\end{figure}
\end{document}